\documentclass[twocolumn,pre,superscriptaddress,10pt,aps,floatfix]{revtex4-1}
\usepackage[english]{babel}
\usepackage{graphicx}
\usepackage{amsmath}
\usepackage[colorlinks,linkcolor=blue,citecolor=blue,urlcolor=blue]{hyperref}
\usepackage{color}
\usepackage{soul}
\definecolor{dblue} {RGB}{28,130,185}
\let\oldaddcontentsline\addcontentsline
\newcommand{\stoptocentries}{\renewcommand{\addcontentsline}[3]{}}
\newcommand{\starttocentries}{\let\addcontentsline\oldaddcontentsline}
\definecolor{nred}{RGB}{224,0,0}
\definecolor{nblue}{RGB}{28,130,185}
\definecolor{darkgreen}{rgb}{0,0.60,.2}

\begin{document}
\title{Strongly disordered Anderson insulator chains with generic two-body interaction}
\author{B. Krajewski}
\affiliation{Institute of Theoretical Physics, Faculty of Fundamental Problems of Technology, \\ Wroc\l aw University of Science and Technology, 50-370 Wroc\l aw, Poland}
\author{L. Vidmar}
\affiliation{Department of Theoretical Physics, J. Stefan Institute, SI-1000 Ljubljana, Slovenia}
\affiliation{Department of Physics, Faculty of Mathematics and Physics, University of Ljubljana, SI-1000 Ljubljana, Slovenia\looseness=-1}
\author{J. Bon\v ca}
\affiliation{Department of Physics, Faculty of Mathematics and Physics, University of Ljubljana, SI-1000 Ljubljana, Slovenia\looseness=-1}
\affiliation{Department of Theoretical Physics, J. Stefan Institute, SI-1000 Ljubljana, Slovenia}
\author{M. Mierzejewski}
\affiliation{Institute of Theoretical Physics, Faculty of Fundamental Problems of Technology, \\ Wroc\l aw University of Science and Technology, 50-370 Wroc\l aw, Poland}
\date{\today}
\begin{abstract}
The random-field spin-1/2 XXZ chains, and the corresponding Anderson insulators of spinless fermions with density-density interaction, have been intensively studied in the context of many-body localization. However, we recently argued [B. Krajewski et al., \href{https://doi.org/10.1103/PhysRevLett.129.260601}{Phys.~Rev.~Lett.~{\bf 129},~260601~(2022)}] that the two-body density-density interaction in these models is not generic since only a small fraction of this interaction represents a true local perturbation to the Anderson insulator.
Here we study ergodicity of strongly disordered Anderson insulator chains choosing other forms of the two-body interaction for which the strength of the true perturbation is of the same order of magnitude as the bare two-body interaction. Focusing on the strong interaction regime, numerical results for the level statistics and the eigenstate thermalization hypothesis are consistent with emergence of ergodicity at arbitrary strong disorder.
  
\end{abstract}
\maketitle
\stoptocentries
%================================================================================

\section{Introduction}

Extensive numerical studies concerning ergodicity of disordered systems have been carried out mostly for the one-dimensional random-field spin-1/2 XXZ model (see, e.g.,~\cite{pal10,barisic10,luitz15,luitz16,torres15,bera15,Hauschild_2016,Devakul2015,bertrand_garcia_16,Doggen2018,suntajs_bonca_20a, suntajs_2020, sierant_lewenstein_20}). 
This spin model can be mapped onto a chain of spinless fermions, for which the Hamiltonian can be expressed as
\begin{equation} \label{def_H_general}
    H = H_0(W) + H'(V) \;,
\end{equation}
where the single-particle term $H_0(W)$ describes the Anderson insulator with the disorder strength $W$ and the hopping amplitude set to unity, while the interaction $H'(V)$ is given by the density-density interaction of strength $V$.  
This is one of the simplest models which allows to study the influence of two-body interaction on the Anderson localization.

At a sufficiently strong disorder, $W>W^*$, numerical studies of finite systems indicated that localization may also persist in the presence of the two-body interaction, leading to the concept of many-body localization (MBL)~\cite{basko06,oganesyan07}. In particular, the presence of MBL has been suggested based on extremely slow dynamics~\cite{znidaric08,serbyn13_1,mierzejewski2016,luschen17,bordia2017_1,serbyn2017,Bera2019,chanda2020}, suppressed transport~\cite{berkelbach10, lev15, barisic16, steinigeweg16, bera2017, prelovsek116}, entanglement entropy properties~\cite{znidaric08,bardarson12,kjall14,baygan15, pietracaprina_parisi_17} and level statistics analyses~\cite{oganesyan07,pal10,luitz15,zakrzewski16,sierant_zakrzewski_19}.
 
In spite of these results, however, the fate of localization in macroscopic systems remains an open problem. 
In particular, the transition/crossover disorder $W^*$, as estimated from the energy spectrum \cite{suntajs_bonca_20a,suntajs_2020}, shows a linear drift with the system size $L$, namely $W^* \propto L +{\rm const}$. The latter observations as well as several other results obtained from subsequent studies suggest that macroscopic systems may be ergodic~\cite{kieferemmanouilidis_unanyan_20, sels2020, kieferemmanouilidis_unanyan_21, leblond_sels_21, vidmar2021, Sels_dilute_2021,Sels_2022,evers2023} and diffusive, although with a very small diffusion constant~\cite{barisic10}. At even larger disorders, much larger than the previous estimates of $W^*$~\cite{sierant2022,Morningstar2022}, the state of the art numerical approaches may not provide a conclusive answer to this question\cite{abanin2019,Panda2020,abanin_bardarson_21,sierant_lewenstein_20,crowley_chandran_22,bolther_kehrein_22}.

A nontrivial feature of the $V-W$ state diagram in finite systems is the emergence of reentrant behavior~\cite{deluca13,lev15}.
This means that at fixed $W$, ergodicity is restored at some nonzero interaction $V_1$, while at some larger interaction $V_2>V_1$ one again observes the breakdown of ergodicity.
This property may suggest that the density-density interaction cannot be considered as a perturbation that just destroys the Anderson localization of noninteracting particles; it may also stabilize localization once it is large enough. 

In a recent Letter~\cite{krajewski_vidmar_22}, we showed that only a small part of the density-density interaction $H'(V)$, termed {\it true local perturbation} (shortly, true perturbation), does not commute with $H_0$. Namely, the major part of $H'(V)$ can be expressed via products of occupations of the single-particle Anderson states, and hence this part of $H'(V)$ does not perturb  the Anderson insulator. 
At strong disorder, the true perturbation may hence be too small to be correctly captured by numerical calculations carried out in finite systems.
By rescaling the true perturbation so that its strength is of the same order as the other terms of the Hamiltonian, we introduced a rescaled model that exhibits ergodicity at arbitrary strong $W$~\cite{krajewski_vidmar_22}. However, this approach is computationally demanding and the spatial structure of the true perturbation is rather complex.  

In this work, we demonstrate that the smallness of the true perturbation is specific for models with the density-density interaction, and that other forms of the two-body interaction do not share this property. 
Therefore,  the emergence and breakdown of ergodicity  in the latter systems can be studied without applying a complex procedure that singles out and rescales  the true perturbation.
As the main result, we show evidence that a sufficiently strong two-body interaction restores ergodicity even at extremely large disorder such as $W\sim 20$.
This observation also suggests the absence of reentrant behavior at fixed disorder $W$ when interaction is increased.

The paper is organized as follows.
In Sec.~\ref{sec:models}, we recall the notion of the true perturbation and introduce the studied forms of the two-body interactions. Then, in Sec.~\ref{results} we discuss numerical results obtained for the level statistics and the eigenstate thermalization hypothesis (ETH)~\cite{deutsch_91,srednicki_94,rigol_dunjko_08,dalessio_kafri_16}.
We summarize and discuss our results in Sec.~\ref{sec:discussion}.
  
\section{Models and the strength of  the true perturbation} \label{sec:models}
 
We consider a chain of length $L$ containing $L/2$  spinless fermions, where we assume periodic boundary conditions. We discuss the ergodicity of three distinct models of the form~(\ref{def_H_general}), which share the same single-particle part, $H_0$, but have different two-body interaction terms, $H'$. In particular, we take
\begin{equation}
H_0=\frac{t_h}{2} \sum_i \left(c^{\dagger}_{i+1} c_i +{\rm H.c.} \right) +\sum_i \epsilon_i \left(n_i-\frac{1}{2}\right), \label{H_0}
\end{equation}
where $c^{\dagger}_i$ creates a fermion at site  $i$, $n_i=c^{\dagger}_i c_i$ and $\epsilon_i$ is a random potential (i.e., the disorder) uniformly distributed within the box $\epsilon_i \in [-W,W]$. From this point on, we set $t_h=1$.
We refer to the model in Eq.~(\ref{H_0}) as the Anderson model.
The noninteracting part is diagonal in the Anderson single-particle basis:
\begin{equation}
H_0= \sum_{\alpha} \frac{1}{2} \varepsilon_{\alpha}  Q_{\alpha} +{\rm const}, \quad \quad   Q_{\alpha}=2 n_{\alpha} -1, \label{H_0_diag}
\end{equation}     
where $n_{\alpha}$ is the particle number operator for the Anderson state $|\alpha \rangle$ with the energy $\varepsilon_{\alpha}$, and $Q_{\alpha}$ are referred to as the Anderson local integrals of motions (shortly, Anderson LIOMs).  It is convenient to single out the operator density, $h(i)$, of the two-body interaction, and express $H'$ as
\begin{equation}
H' = V \sum_i h(i)\;. \label{h'}
\end{equation}
The first model we have in mind is the standard model of MBL with the nearest-neighbor density-density (dd) interaction for spinless fermions,
\begin{equation}
h_{\rm dd}(i)  =  (n_i-\tfrac{1}{2})(n_{i+1}-\tfrac{1}{2})\;, \label{h'dd}
\end{equation}
which maps onto the $S^z_i S^z_{i+1}$ interaction in the random-field spin-1/2 XXZ model.
Increasing the disorder in the Anderson model in Eq.~(\ref{H_0}) yields a shorter localization length, and ultimately at large $W$ each single-particle Anderson wave function $\langle i | \alpha \rangle$ is strongly peaked at a single site $i=i(\alpha)$.
Then, the occupations in Eq.~(\ref{h'dd}) become {\em similar} to the Anderson LIOMs in Eq.~(\ref{H_0_diag}), \mbox{$ n_{i(\alpha)}-\frac{1}{2} \simeq \frac{1}{2} Q_{\alpha}$}. Therefore, the value of $V$ alone does not specify whether  \mbox{$H'_{\rm dd} = V \sum_i h_{\rm dd}(i)$}  
represents strong or weak perturbation to $H_0$. 

In order to quantify this intuitive understanding [especially the {\em similarity} between $n_{i(\alpha)}-\frac{1}{2}$ and $Q_{\alpha}$], we use the Hilbert-Schmidt inner product of two operators $\langle A  B \rangle=\frac{1}{Z} {\rm Tr}(A^\dagger  B)$ and the corresponding norm of operators $||A||^2=\langle A A \rangle$.  Here, the trace is carried out over the many-body states and $Z = {\binom{L}{L/2}}$ is the dimension of the many-body Hilbert space. 
We split the perturbation operators from Eq.~(\ref{h'dd}) into two parts, 
\begin{equation} \label{def_hdd_parts}
h_{\rm dd}(i)=h^{\parallel}_{\rm dd}(i)+h^{\perp}_{\rm dd}(i)\;,
\end{equation}
in such a way that both parts are mutually orthogonal 
$\langle h^{\parallel}_{\rm dd}(i) h^{\perp}_{\rm dd}(i) \rangle=0$ and $h^{\parallel}_{\rm dd}(i)$ commutes with $H_0$. Therefore, only the latter part,  $V h^{\perp}_{\rm dd}(i)$, represents the density of the true local perturbation, and the operators can be expressed as (see~\cite{krajewski_vidmar_22} for details)
\begin{eqnarray}
h^{\parallel}_{\rm dd}(i)&=&\sum_{\alpha,d \le d_{\rm max}} \langle  h_{\rm dd}(i)  Q_{\alpha}Q_{\alpha+d}  \rangle  \;\; Q_{\alpha}Q_{\alpha+d}\;, \label{hpar} \\
h^{\perp}_{\rm dd}(i)&=& h_{\rm dd}(i)  - h^{\parallel}_{\rm dd}(i)\;. \label{hperp} 
\end{eqnarray} 
The index $\alpha+d$ of $Q_{\alpha+d}$ in Eq.~(\ref{hpar}) corresponds to the single-particle Anderson state for which the maximum of the wave function $|\langle i | \alpha+d \rangle|$ is shifted left by $d$ sites with respect to the maximum of $|\langle i | \alpha \rangle|$.
We recall that at strong disorder the strength of the true perturbation decays with $W$ as $V|| h^{\perp}_{\rm dd}(i) || \sim  V/W$, see Fig.~1 in~\cite{krajewski_vidmar_22}, whereas $V   h^{\parallel}_{\rm dd}(i)  \sim  V$ should be considered as a part
of the unperturbed energy density.  

As a consequence of the above analysis, the strongly disordered XXZ models always represent weakly perturbed Anderson insulators, independently of $V$. 
Weak perturbations are challenging for numerical calculations which are carried out in finite systems. To avoid possible numerical artifacts, in Ref.~\cite{krajewski_vidmar_22} we rescaled the strength of the true perturbation in such a way that its norm is of the same order of magnitude as the norm of $H_0$. Although the rescaled model was shown to be ergodic, the explicit form of the rescaled Hamiltonian is rather complex and may raise concerns whether the conclusions concerning its ergodicity are generic.
 
To avoid difficulties originating from the presence of a very weak perturbation, in this work we study simple Hamiltonians in which the norms of the single-particle term and the true two-body perturbation are controlled by two independent parameters $W$ and $V$, respectively. 
Since the smallness of the true perturbation in the XXZ chains is specific for the density-density interaction, in this work, we study systems with two-body interactions that cannot be expressed in terms of occupations of the lattice sites.   
We mostly focus on the two-body interaction $h(i)$ from Eq.~(\ref{h'}), which is given by a pair-hopping (ph) term
\begin{equation}
h_{\rm ph}(i)  =  c^{\dagger}_{i-1}  c^{\dagger}_{i} c_{i+1} c_{i+2} +{\rm H.c.}\;\;\; .  \label{ph}
\end{equation}
In order to demonstrate that our conclusions do not originate from a specific choice of the perturbation,  at the end of the paper we also discuss numerical results obtained for the operator
\begin{equation}
\tilde{h}_{\rm ph}(i)  =  c^{\dagger}_{i} c^{\dagger}_{i+1} c_{i+2}  c_{i-1} +{\rm H.c.} \;\;\;  . \label{ph2}
\end{equation}
Results obtained for both models are qualitatively the same. Namely, we show that even at a very strong disorder $W$, a sufficiently strong perturbation restores the ergodicity of the studied model. The larger the system, the weaker the perturbation for which the ergodicity is restored. 

\section{Numerical results} \label{results}

\begin{figure}[!tb]
\includegraphics[width=1.0\columnwidth]{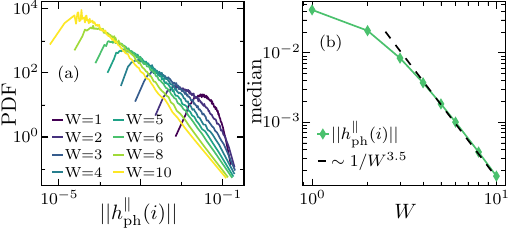}
\caption{Norms of the operators $||h_{\rm ph}^\parallel(i)||$ defined via Eq.~(\ref{hpar}) by replacing $h_{\rm dd}(i) \to h_{\rm ph}(i)$. 
(a) Probability density functions (PDFs) of $||h_{\rm ph}^\parallel(i)||$ at $L=14$ and $d_{\rm max}=2$ for various $W$. The distributions are calculated at all sites for 2000 realizations of the disorder. 
(b) \mbox{$W$-dependence} of the medians of the distributions shown in (a).
The dashed line is the function $\propto 1/W^{3.5}$.
}
\label{fig1}
\end{figure}  

\subsection{The strength of the true perturbation}

We now focus on the Anderson insulators with pair-hopping interaction,
\begin{equation}
H=H_0 + H'(V) = H_0+V\sum_i h_{\rm ph}(i), \label{ham}
\end{equation}
where $h_{\rm ph}(i)$ is defined in Eq.~(\ref{ph}).
First, we demonstrate that the strength of the true perturbation in this model only weakly depends on $W$ and is of the same order as the strength of the bare two-body interaction, $V || h_{\rm ph}(i)||$. To this end, we replace $h_{\rm dd}(i) \to h_{\rm ph}(i)$ in Eq.~(\ref{def_hdd_parts}) and we use the projections defined via Eqs.~(\ref{hpar}) and~(\ref{hperp}) to calculate $ h^{\parallel}_{\rm ph}(i)$ and $ h^{\perp}_{\rm ph}(i)$, respectively. Due to the orthogonality of these two terms one obtains the following identity:
\begin{equation}
|| h_{\rm ph}(i)||^2=||h^{\parallel}_{\rm ph}(i)||^2+ || h^{\perp}_{\rm ph}(i)||^2, \label{norms}
\end{equation}
where the norm on the left-hand side of Eq.~(\ref{norms}) does not depend on $W$ or $V$. Therefore, the strength of the true perturbation, $V|| h^{\perp}_{\rm ph}(i)||$,  can be estimated indirectly from the component which commutes with $H_0$, i.e., from $h^{\parallel}_{\rm ph}(i)$. 

Using exact diagonalization for a system with $L=14$ sites, we calculated the norms of $||h^{\parallel}_{\rm ph}(i)||$ for various realizations of disorder and various sites $i$. From the latter we determined the probability density functions (PDFs) for $||h^{\parallel}_{\rm ph}(i)||$, which are shown in Fig. \ref{fig1}(a) for various $W$. 
In Fig.~\ref{fig1}(b) we show medians of these distributions as a function of the disorder strength $W$. Upon increasing $W$, the maxima of these
distributions shift towards smaller values of $||h^{\parallel}_{\rm ph}(i)||$. Moreover, the medians decrease faster than $1/W^3$.    
This effect is opposite to the results for the density-density interactions, where $||h^{\parallel}_{\rm dd}(i)||$ are very close to $|||h_{\rm dd}(i)||$, whereas
$|||h^{\perp}_{\rm dd}(i)||$ decrease as $1/W$. In the present case
$||h^{\parallel}_{\rm ph}(i)||$ are negligible, in particular at strong disorder. 

To summarize, based on the results in Fig.~\ref{fig1} and using the identity from Eq.~$(\ref{norms})$, we conclude that the strength of the true perturbation in the model (\ref{ham}) is determined mainly by $V$ and, up to reasonable accuracy, one may assume that it is equal to the bare interaction $H'$.

\subsection{Average level spacing ratio}

To identify the ergodic and nonergodic regimes of the model from Eq.~(\ref{ham}), we start with the commonly studied ergodicity indicator, i.e., the ratio  $r_n={\rm min}(\delta E_n,\delta E_{n+1})/{\rm max}(\delta E_n,\delta E_{n+1})$   of  the nearest level spacings $\delta E_n=E_{n+1}-E_{n}$, where $E_n$ are (sorted) energy levels~\cite{oganesyan07, atas2013}. 
We average $r_n$ over the middle third of the energy spectrum, as well as over $4000$ realizations of the disorder, and we denote the average level spacing ratio as $\langle r \rangle$.

\begin{figure}[!tb]
\includegraphics[width=1.0\columnwidth]{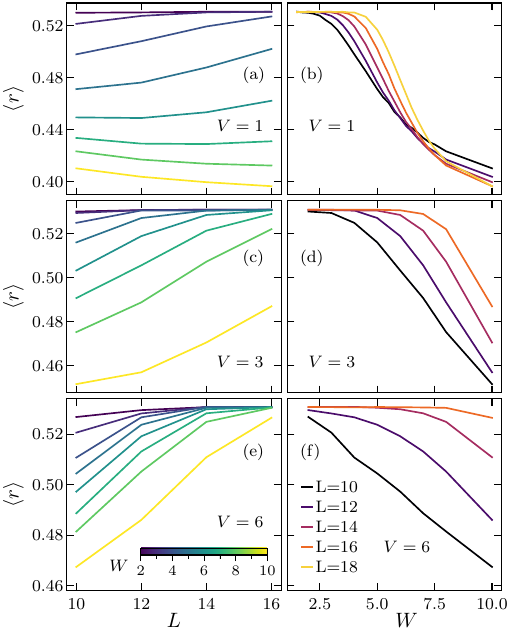}
\caption{Average level spacing ratio $\langle r\rangle$ for the model in Eq.~(\ref{ham}). The averaging is carried out over $Z/3$ levels from the middle of the spectrum and over 4000 realizations of disorder. 
(a,c,e) Dependence of $\langle r\rangle$ on $L$ at interactions $V=1,3,6$, respectively.
Dark to light colors denote the disorders $W=2,3,4,5,6,7,8,10$, see also the colorbar in (e).
(b,d,f) Dependence of $\langle r\rangle$ on $W$ at interactions $V=1,3,6$, respectively.
Dark to light colors denote the system sizes $L=10,12,14,16,18$, see also the legend in (f). 
}
\label{fig2}
\end{figure}

Figures~\ref{fig2}(a), \ref{fig2}(c) and \ref{fig2}(e) show results for $\langle r \rangle$ versus $L$ at $V=1,3,6$, respectively, obtained at $W=2,3,4,5,6,7,8,10$.
In contrast, Figs.~\ref{fig2}(b), \ref{fig2}(d) and \ref{fig2}(f) show results for $\langle r \rangle$ versus $W$ and the same $V=1,3,6$, respectively, obtained at $L=10,12,14,16,18$.

\begin{figure}[!tb]
\includegraphics[width=1.0\columnwidth]{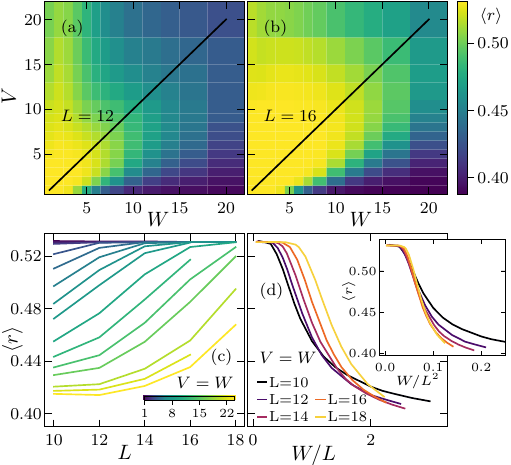}
\caption{Average level spacing ratio $\langle r\rangle$ for the model in Eq.~(\ref{ham}).
(a,b) Heat maps of $\langle r \rangle$ at $L=12$ and 16, respectively.
The solid line denotes $V=W$.
(c) Dependence of $\langle r\rangle$ on $L$  at $V=W$.
Dark to light colors denote the disorders $W=1,2,3,4,5,6,7,8,10,12,14,16,20,22,24$, see also the colorbar.
(d) Dependence of $\langle r \rangle$ on $W/L$ (main panel) and  $W/L^2$ (inset) at $V=W$.
Dark to light colors denote the system sizes $L=10,12,14,16,18$, see also the legend.
}
\label{fig3}
\end{figure}  

The results at weak perturbation $V=1$, see Figs.~\ref{fig2}(a) and~\ref{fig2}(b), share some similarities with the numerical results obtained previously for disordered chains with density-density interaction~\cite{luitz15,bertrand_garcia_16,Husex2017, suntajs_2020, sierant_lewenstein_20}.
In particular, at sufficiently large disorder, see Fig.~\ref{fig2}(a), $\langle r \rangle$ decreases with $L$ towards the value for Poisson level statistics $r_{\rm Pois}\simeq 0.39$, eventually indicating nonergodicity.
The finite-size phenomenology of the results for $\langle r \rangle$ in systems with density-density interaction was extensively studied in the past, giving rise to conjectures either about robustness of ergodicity~\cite{suntajs_2020} or about the convergence to a critical point~\cite{sierant_lewenstein_20}.
Our study does not contribute significantly to new results in this regard.

In contrast to systems with density-density interaction, in the present model~(\ref{ham}) one may tune the strength of the interaction and explore a broader regime of parameters where the numerical results do not suffer from suppression of the true perturbation.  
Figures~\ref{fig2}(c)-\ref{fig2}(f) show results for $\langle r \rangle$ at stronger interactions $V=3$ and 6. 
They show a strong tendency towards restoring ergodicity.
For example, in the system with $L=16$ sites, it is reasonable to expect $\langle r \rangle$ to approach Gaussian orthogonal ensemble (GOE) prediction $r_{\rm GOE}\simeq 0.53$ at disorders of strengths $W \approx 10$.
Moreover, the flow of the results with increasing $L$ suggests restoration of ergodicity also at disorders much larger than $W \approx 10$.

A different perspective of the finite-size phenomenology of $\langle r \rangle$ is presented in Fig.~\ref{fig3}.
The values of $\langle r \rangle$ in the $V-W$ plane of parameters are shown as heat maps in Figs.~\ref{fig3}(a) and~\ref{fig3}(b) at $L=12$ and 16, respectively.
In both cases, signatures of a reentrant behavior may be observed, i.e., at a fixed $W$ and by increasing $V$, $\langle r \rangle$ first increases towards $r_{\rm GOE}$ and then decreases again.
However, another important observation is also that by increasing the system size from $L=12$ to $L=16$, there is a strong enhancement of the region in which $r \approx r_{\rm GOE}$.

To study the fate of the reentrant behavior when increasing the system size, we focus on the parameter line $V=W$, which is shown as a solid line in Figs.~\ref{fig3}(a) and~\ref{fig3}(b).
If the system exhibits a tendency towards restoring ergodicity along this line, it is unlikely that the reentrant behavior survives in the thermodynamic limit.

Figure~\ref{fig3}(c) shows results for $\langle r \rangle$ versus $L$ at $V=W$.
Even for the largest disorder under consideration, $W =24$, we clearly observe a drift of $\langle r  \rangle $ towards $r_{\rm GOE}$ upon increasing $L$.
A different view of these results can be obtained from Fig.~\ref{fig3}(d), in which we show $\langle r \rangle$ versus $W/L$ at $V=W$.
If the characteristic disorder at which $\langle r \rangle$ starts to depart from $r_{\rm GOE}$ drifts linearly with $L$, as in the case of density-density interactions~\cite{suntajs_2020}, one should observe a collapse of curves for $\langle r \rangle$ at different $L$ when plotted versus $W/L$.
However, the results in Fig.~\ref{fig3}(d) show signatures that the drift is faster than linear. In particular, the values of disorder at which
$\langle r  \rangle $ starts to deviate from $r_{\rm GOE}$ seem to increase with system size as $W^* \propto L^2$, see the inset in Fig.~\ref{fig3}(d). 
The latter scaling is different from the scaling $W^*\propto L$ at $V=W$ line for the density-density interaction (not shown).
We interpret this result as a strong evidence that the system in Eq. (\ref{ham})
 becomes ergodic along the $V=W$ parameter line in the thermodynamic limit; hence the state diagram is not expected to exhibit reentrant behavior.

\subsection{Diagonal matrix elements of observables and the ETH} \label{sec:ETH}

Next, we demonstrate the tendency of the studied model~(\ref{ham}) to become ergodic via testing the ETH. We perform numerical calculations for the site occupation operator $A_i=2n_i-1$.  This observable (or some linear combinations of $A_i$) was commonly studied in the context of MBL~\cite{mierzejewski2016,luitz16,sirker14,pal10} because localization would imply that the matrix elements of $A_i$ must violate the ETH. The latter violation would show up in the nonvanishing fluctuations of the diagonal matrix elements $\langle E_m| A_i |E_m \rangle $, which determine the value of the infinite-time correlation function $\lim_{t\to \infty} \langle A_i(t) A_i \rangle$. 

In order to estimate the fluctuations of the diagonal matrix elements, we calculate the average eigenstate-to-eigenstate fluctuations~\cite{Kim_strong2014, Mondaini2016, jansen_stolpp_19},
\begin{equation} \label{eefluct}
\langle \delta A_i \rangle= \frac{1}{\cal{Z}} \sum_{m} |\langle E_{m+1}| A_i |E_{m+1} \rangle -\langle E_m| A_i |E_m \rangle | \;,
\end{equation}
where the eigenstate-to-eigenstate fluctuations are averaged over ${\cal{Z}}=Z/5$ eigenstates from the middle of the many-body spectrum. 

\begin{figure}[!tb]
\includegraphics[width=1.0\columnwidth]{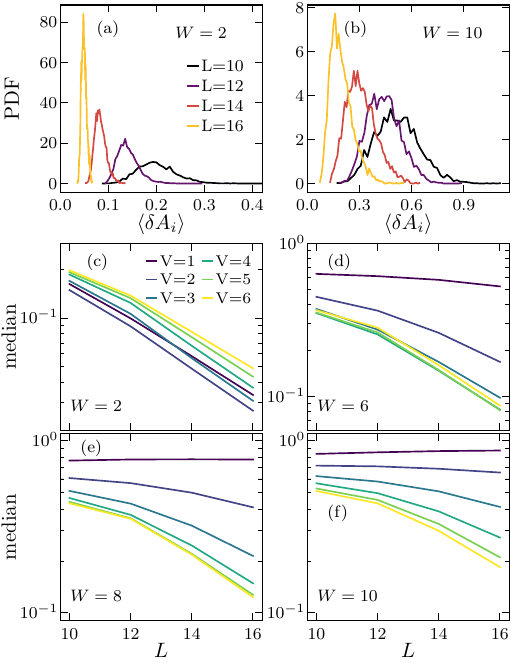}
\caption{Fluctuations of the diagonal matrix elements for the model defined in Eq.~(\ref{ham}). (a) and (b): PDFs of $\langle\delta A_i\rangle$, see  Eq.~(\ref{eefluct}), 
at $V=6$ for a single site $i$ and 4000 realizations of disorder.  (c)-(f): Medians of distributions as
in (a) and (b) versus $L$ for various $V$ and $W$.}
\label{fig4}
\end{figure}

Figures~\ref{fig4}(a) and~\ref{fig4}(b) show the PDFs of $\langle \delta A_i \rangle$ obtained from a collection of results for 4000 realizations of disorder. In the case of a finite system, $A_i$ at different sites are not independent since $\sum_i A_i=0$. Therefore, we calculate $A_i$ for a single lattice site $i$. These PDFs were obtained at $V=6$, when the average level spacing ratio $\langle r \rangle$ matches the GOE prediction.
At strong disorder $W=10$, see Fig.~\ref{fig4}(b), the values of $\langle \delta A_i \rangle$ remain large and the presented distributions are broad. The latter feature indicates the presence of substantial sample-to-sample fluctuation, which seems to be unavoidable in small systems at strong disorder~\cite{mierzejewski2020, krajewski2022}. Nevertheless, upon increasing $L$ the distributions become narrower and their maxima shift towards smaller eigenstate-to-eigenstate fluctuations. This indicates that the diagonal matrix elements
of $A_i$ satisfy the ETH. 

In order to study in more detail how the distributions of $\langle \delta A_i \rangle$ depend on $V$ and $W$, in Figs.~\ref{fig4}(c)-\ref{fig4}(f) we show the
medians of these distributions versus $L$. The conclusions are quite similar to those derived previously from the level statistics. 
At weak two-body interactions, see the results for $V=1$ in Figs.~\ref{fig4}(e) and~\ref{fig4}(f), strongly disordered systems seem to violate the ETH since the eigenstate-to-eigenstate fluctuations stay large for all accessible system sizes. However, the $L$-dependence of the medians at $V>1$ exhibits clear negative curvatures in this regime of parameters, suggesting that
the eigenstate-to-eigenstate fluctuations should eventually start decaying for sufficiently large systems. In the case of stronger two-body interaction, which is the main focus of our study, such decay is clearly visible for all studied disorder strengths. 

\subsection{Results for two-body interaction from Eq.~(\ref{ph2})}

\begin{figure}[!tb]
\includegraphics[width=1.0\columnwidth]{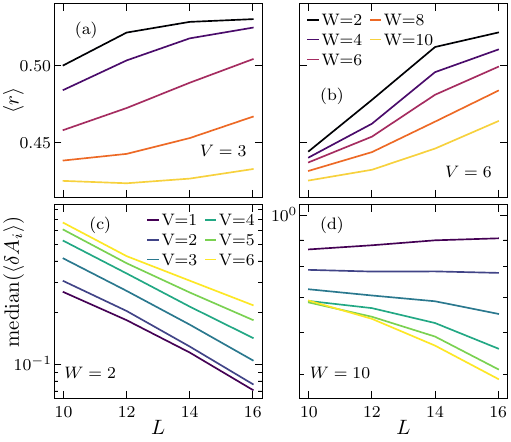}
\caption{Ergodicity indicators for the model defined in Eq.~(\ref{ham2}). (a) and (b): Average level spacing ratio $\langle r \rangle$ versus $L$ at $V=3$ and 6, respectively.
Results are analogous to those in Figs.~\ref{fig2}(c) and \ref{fig2}(e), respectively, shown for the model from Eq.~(\ref{ham}).
(c) and (d) Medians of the PDFs of the eigenstate-to-eigenstate fluctuations $\langle \delta A_i \rangle$ from Eq.~(\ref{eefluct}) at $W=2$ and 10, respectively.
Results are analogous to those in Figs.~\ref{fig4}(c) and \ref{fig4}(f), respectively, shown for the model from Eq.~(\ref{ham}).}
\label{fig5}
\end{figure}

To demonstrate that the ergodicity of the model studied so far is not restricted to the particular form of the interaction, in this section we discuss the average level spacing ratio and the fluctuations of the diagonal matrix elements of observables for the Hamiltonian
\begin{equation}
\tilde{H}=H_0+V\sum_i \tilde{h}_{\rm ph}(i), \label{ham2}
\end{equation}
where the pair-hopping term, $\tilde{h}_{\rm ph}(i)$, is given by Eq. (\ref{ph2}). Similarly to the previously discussed case, also $||\tilde{h}^{\parallel}_{\rm ph}(i)||$ decays with increasing $W$ (not shown); therefore, the entire two-body interaction can be considered as a perturbation to $H_0$. 

The average level spacing ratios $\langle r \rangle$ obtained at weak interaction $V=1$ (not shown) are very similar to the results for the previously studied model from Eq.~(\ref{ham}) shown in 
Figs.~\ref{fig2}(a) and \ref{fig2}(b). The results at stronger interactions $V=3$ and 6 are shown in Figs.~\ref{fig5}(a) and~\ref{fig5}(b), respectively. Although deviations of $\langle r \rangle$ from $r_{\rm GOE}$ are more pronounced than in the case of the model from Eq.~(\ref{ham}),
the $L$-dependence of  $\langle r \rangle $ consistently supports the expectation that the spectral properties of macroscopic systems should approach the GOE predictions.

Finally, we also calculated the fluctuations of the diagonal matrix elements $\langle \delta A_i \rangle$ from Eq.~(\ref{eefluct}). Similarly to the discussion in the Sec.~\ref{sec:ETH},
we obtained the PDFs of $\langle\delta A_i \rangle$, and in Figs.~\ref{fig5}(c) and  \ref{fig5}(d) we show the corresponding medians. Accessible system sizes do not allow one to formulate univocal claims concerning ergodicity of macroscopic systems at very strong disorder and weak two-body interactions. However, results in the strong interaction regime clearly indicate that the system satisfies the ETH.

\section{Discussion} \label{sec:discussion}

We studied the ergodicity of Anderson insulator chains with two-body interactions. A similar problem has been widely studied for systems with nearest-neighbor density-density interactions, mappable onto the random-field spin-1/2 XXZ chain. Due to its simplicity, the random-field XXZ chain appears to be the most natural model for the studies on MBL. However, the interaction term in the fermionic models mappable onto the XXZ chains is rather specific in that it has a large projection on (products) of LIOMs of the single-particle Anderson insulator. Due to this important property, 
only a tiny part of the two-body interaction represents a true perturbation to the Anderson insulators. 

One may expect that a generic local two-body interaction also contains other terms which have no projections on the Anderson LIOMs and thus have greater impact on the ergodicity of disordered systems. For this reason, in the present work we have studied two cases of the pair-hopping interactions which have negligible projections on the Anderson LIOMs.

We found that the results for finite systems at strong disorder $W$ and weak two-body interaction $V$ qualitatively reproduce the known results for random-field XXZ chains.
However, the regime with exceedingly different model parameters is most demanding for numerical calculations. 
For this reason, our results neither confirm nor contradict the presence of localization for large $W$ and $V/W \ll 1$.   
In the opposite case where $V$ and $W$ are of the same order of magnitude, which was the main focus of this paper, our numerical results indicate that macroscopic systems are ergodic. 
The latter observation follows from our studies of the average level spacing ratios as well as the fluctuations of the diagonal matrix elements of local observables.

\acknowledgements 
We acknowledge support from the National Science Centre, Poland via Project No. 2020/37/B/ST3/00020 (B.K and M.M.)and the support from the Slovenian Research Agency (ARRS), Research Core Funding Grants No. P1-0044 (L.V. and J.B.) and No. N1-0273 (L.V.). The numerical calculation were partly carried out at the facilities of
the Wrocław Centre for Networking and Supercomputing.
 
\bibliographystyle{biblev1}
\bibliography{ref_mbl}

\begin{thebibliography}{10}
\expandafter\ifx\csname url\endcsname\relax
  \def\url#1{{\tt #1}}\fi
\expandafter\ifx\csname urlprefix\endcsname\relax\def\urlprefix{URL }\fi
\expandafter\ifx\csname bibinfo\endcsname\relax\def\bibinfo#1#2{#2}\fi
\expandafter\ifx\csname eprint\endcsname\relax\def\eprint#1{\url{#1}}\fi

\bibitem{pal10}
\bibinfo{author}{A.~Pal} and \bibinfo{author}{D.~A. Huse},
  \bibinfo{title}{Many-body localization phase transition},
  \bibinfo{journal}{\href{http://dx.doi.org/10.1103/PhysRevB.82.174411}{Phys.
  Rev. B}} \href{http://dx.doi.org/10.1103/PhysRevB.82.174411}{{\bf
  \bibinfo{volume}{82}}, \bibinfo{pages}{174411}}
  (\href{http://dx.doi.org/10.1103/PhysRevB.82.174411}{\bibinfo{year}{2010}}).

\bibitem{barisic10}
\bibinfo{author}{O.~S. Bari\ifmmode \check{s}\else
  \v{s}\fi{}i\ifmmode~\acute{c}\else \'{c}\fi{}} and
  \bibinfo{author}{P.~Prelov\ifmmode~\check{s}\else \v{s}\fi{}ek},
  \bibinfo{title}{Conductivity in a disordered one-dimensional system of
  interacting fermions},
  \bibinfo{journal}{\href{http://dx.doi.org/10.1103/PhysRevB.82.161106}{Phys.
  Rev. B}} \href{http://dx.doi.org/10.1103/PhysRevB.82.161106}{{\bf
  \bibinfo{volume}{82}}, \bibinfo{pages}{161106}}
  (\href{http://dx.doi.org/10.1103/PhysRevB.82.161106}{\bibinfo{year}{2010}}).

\bibitem{luitz15}
\bibinfo{author}{D.~J. Luitz}, \bibinfo{author}{N.~Laflorencie}, and
  \bibinfo{author}{F.~Alet}, \bibinfo{title}{Many-body localization edge in the
  random-field {Heisenberg} chain},
  \bibinfo{journal}{\href{http://dx.doi.org/10.1103/PhysRevB.91.081103}{Phys.
  Rev. B}} \href{http://dx.doi.org/10.1103/PhysRevB.91.081103}{{\bf
  \bibinfo{volume}{91}}, \bibinfo{pages}{081103}}
  (\href{http://dx.doi.org/10.1103/PhysRevB.91.081103}{\bibinfo{year}{2015}}).

\bibitem{luitz16}
\bibinfo{author}{D.~J. Luitz}, \bibinfo{author}{N.~Laflorencie}, and
  \bibinfo{author}{F.~Alet}, \bibinfo{title}{{Extended slow dynamical regime
  prefiguring the many-body localization transition}},
  \bibinfo{journal}{\href{http://dx.doi.org/10.1103/PhysRevB.93.060201}{Phys.
  Rev. B}} \href{http://dx.doi.org/10.1103/PhysRevB.93.060201}{{\bf
  \bibinfo{volume}{93}}, \bibinfo{pages}{060201(R)}}
  (\href{http://dx.doi.org/10.1103/PhysRevB.93.060201}{\bibinfo{year}{2016}}).

\bibitem{torres15}
\bibinfo{author}{E.~J. Torres-Herrera} and \bibinfo{author}{L.~F. Santos},
  \bibinfo{title}{Dynamics at the many-body localization transition},
  \bibinfo{journal}{\href{http://dx.doi.org/10.1103/PhysRevB.92.014208}{Phys.
  Rev. B}} \href{http://dx.doi.org/10.1103/PhysRevB.92.014208}{{\bf
  \bibinfo{volume}{92}}, \bibinfo{pages}{014208}}
  (\href{http://dx.doi.org/10.1103/PhysRevB.92.014208}{\bibinfo{year}{2015}}).

\bibitem{bera15}
\bibinfo{author}{S.~Bera}, \bibinfo{author}{H.~Schomerus},
  \bibinfo{author}{F.~Heidrich-Meisner}, and \bibinfo{author}{J.~H. Bardarson},
  \bibinfo{title}{Many-body localization characterized from a one-particle
  perspective},
  \bibinfo{journal}{\href{http://dx.doi.org/10.1103/PhysRevLett.115.046603}{Phys.
  Rev. Lett.}} \href{http://dx.doi.org/10.1103/PhysRevLett.115.046603}{{\bf
  \bibinfo{volume}{115}}, \bibinfo{pages}{046603}}
  (\href{http://dx.doi.org/10.1103/PhysRevLett.115.046603}{\bibinfo{year}{2015}}).

\bibitem{Hauschild_2016}
\bibinfo{author}{J.~Hauschild}, \bibinfo{author}{F.~Heidrich-Meisner}, and
  \bibinfo{author}{F.~Pollmann}, \bibinfo{title}{Domain-wall melting as a probe
  of many-body localization},
  \bibinfo{journal}{\href{http://dx.doi.org/10.1103/physrevb.94.161109}{Physical
  Review B}} \href{http://dx.doi.org/10.1103/physrevb.94.161109}{{\bf
  \bibinfo{volume}{94}}, \bibinfo{pages}{161109}}
  (\href{http://dx.doi.org/10.1103/physrevb.94.161109}{\bibinfo{year}{2016}}).

\bibitem{Devakul2015}
\bibinfo{author}{T.~Devakul} and \bibinfo{author}{R.~R.~P. Singh},
  \bibinfo{title}{Early breakdown of area-law entanglement at the many-body
  delocalization transition},
  \bibinfo{journal}{\href{http://dx.doi.org/10.1103/PhysRevLett.115.187201}{Phys.
  Rev. Lett.}} \href{http://dx.doi.org/10.1103/PhysRevLett.115.187201}{{\bf
  \bibinfo{volume}{115}}, \bibinfo{pages}{187201}}
  (\href{http://dx.doi.org/10.1103/PhysRevLett.115.187201}{\bibinfo{year}{2015}}).

\bibitem{bertrand_garcia_16}
\bibinfo{author}{C.~L. Bertrand} and \bibinfo{author}{A.~M.
  Garc\'{\i}a-Garc\'{\i}a}, \bibinfo{title}{{Anomalous Thouless energy and
  critical statistics on the metallic side of the many-body localization
  transition}},
  \bibinfo{journal}{\href{http://dx.doi.org/10.1103/PhysRevB.94.144201}{Phys.
  Rev. B}} \href{http://dx.doi.org/10.1103/PhysRevB.94.144201}{{\bf
  \bibinfo{volume}{94}}, \bibinfo{pages}{144201}}
  (\href{http://dx.doi.org/10.1103/PhysRevB.94.144201}{\bibinfo{year}{2016}}).

\bibitem{Doggen2018}
\bibinfo{author}{E.~V.~H. Doggen}, \bibinfo{author}{F.~Schindler},
  \bibinfo{author}{K.~S. Tikhonov}, \bibinfo{author}{A.~D. Mirlin},
  \bibinfo{author}{T.~Neupert}, \bibinfo{author}{D.~G. Polyakov}, and
  \bibinfo{author}{I.~V. Gornyi}, \bibinfo{title}{Many-body localization and
  delocalization in large quantum chains},
  \bibinfo{journal}{\href{http://dx.doi.org/10.1103/PhysRevB.98.174202}{Phys.
  Rev. B}} \href{http://dx.doi.org/10.1103/PhysRevB.98.174202}{{\bf
  \bibinfo{volume}{98}}, \bibinfo{pages}{174202}}
  (\href{http://dx.doi.org/10.1103/PhysRevB.98.174202}{\bibinfo{year}{2018}}).

\bibitem{suntajs_bonca_20a}
\bibinfo{author}{J.~\v{S}untajs}, \bibinfo{author}{J.~Bon\v{c}a},
  \bibinfo{author}{T.~Prosen}, and \bibinfo{author}{L.~Vidmar},
  \bibinfo{title}{Quantum chaos challenges many-body localization},
  \bibinfo{journal}{\href{http://dx.doi.org/10.1103/PhysRevE.102.062144}{Phys.
  Rev. E}} \href{http://dx.doi.org/10.1103/PhysRevE.102.062144}{{\bf
  \bibinfo{volume}{102}}, \bibinfo{pages}{062144}}
  (\href{http://dx.doi.org/10.1103/PhysRevE.102.062144}{\bibinfo{year}{2020}}).

\bibitem{suntajs_2020}
\bibinfo{author}{J.~\v{S}untajs}, \bibinfo{author}{J.~Bon\v{c}a},
  \bibinfo{author}{T.~Prosen}, and \bibinfo{author}{L.~Vidmar},
  \bibinfo{title}{Ergodicity breaking transition in finite disordered spin
  chains},
  \bibinfo{journal}{\href{http://dx.doi.org/10.1103/PhysRevB.102.064207}{Phys.
  Rev. B}} \href{http://dx.doi.org/10.1103/PhysRevB.102.064207}{{\bf
  \bibinfo{volume}{102}}, \bibinfo{pages}{064207}}
  (\href{http://dx.doi.org/10.1103/PhysRevB.102.064207}{\bibinfo{year}{2020}}).

\bibitem{sierant_lewenstein_20}
\bibinfo{author}{P.~Sierant}, \bibinfo{author}{M.~Lewenstein}, and
  \bibinfo{author}{J.~Zakrzewski}, \bibinfo{title}{Polynomially filtered exact
  diagonalization approach to many-body localization},
  \bibinfo{journal}{\href{http://dx.doi.org/10.1103/PhysRevLett.125.156601}{Phys.
  Rev. Lett.}} \href{http://dx.doi.org/10.1103/PhysRevLett.125.156601}{{\bf
  \bibinfo{volume}{125}}, \bibinfo{pages}{156601}}
  (\href{http://dx.doi.org/10.1103/PhysRevLett.125.156601}{\bibinfo{year}{2020}}).

\bibitem{basko06}
\bibinfo{author}{D.~Basko}, \bibinfo{author}{I.~Aleiner}, and
  \bibinfo{author}{B.~Altshuler}, \bibinfo{title}{Metal--insulator transition
  in a weakly interacting many-electron system with localized single-particle
  states},
  \bibinfo{journal}{\href{http://dx.doi.org/10.1016/j.aop.2005.11.014}{Ann.
  Phys.}} \href{http://dx.doi.org/10.1016/j.aop.2005.11.014}{{\bf
  \bibinfo{volume}{321}}, \bibinfo{pages}{1126}}
  (\href{http://dx.doi.org/10.1016/j.aop.2005.11.014}{\bibinfo{year}{2006}}).

\bibitem{oganesyan07}
\bibinfo{author}{V.~Oganesyan} and \bibinfo{author}{D.~A. Huse},
  \bibinfo{title}{Localization of interacting fermions at high temperature},
  \bibinfo{journal}{\href{http://dx.doi.org/10.1103/PhysRevB.75.155111}{Phys.
  Rev. B}} \href{http://dx.doi.org/10.1103/PhysRevB.75.155111}{{\bf
  \bibinfo{volume}{75}}, \bibinfo{pages}{155111}}
  (\href{http://dx.doi.org/10.1103/PhysRevB.75.155111}{\bibinfo{year}{2007}}).

\bibitem{znidaric08}
\bibinfo{author}{M.~\v{Z}nidari\v{c}}, \bibinfo{author}{T.~Prosen}, and
  \bibinfo{author}{P.~Prelov\ifmmode~\check{s}\else \v{s}\fi{}ek},
  \bibinfo{title}{Many-body localization in the {Heisenberg XXZ} magnet in a
  random field},
  \bibinfo{journal}{\href{http://dx.doi.org/10.1103/PhysRevB.77.064426}{Phys.
  Rev. B}} \href{http://dx.doi.org/10.1103/PhysRevB.77.064426}{{\bf
  \bibinfo{volume}{77}}, \bibinfo{pages}{064426}}
  (\href{http://dx.doi.org/10.1103/PhysRevB.77.064426}{\bibinfo{year}{2008}}).

\bibitem{serbyn13_1}
\bibinfo{author}{M.~Serbyn}, \bibinfo{author}{Z.~Papi\'{c}}, and
  \bibinfo{author}{D.~A. Abanin}, \bibinfo{title}{Universal slow growth of
  entanglement in interacting strongly disordered systems},
  \bibinfo{journal}{\href{http://dx.doi.org/10.1103/PhysRevLett.110.260601}{Phys.
  Rev. Lett.}} \href{http://dx.doi.org/10.1103/PhysRevLett.110.260601}{{\bf
  \bibinfo{volume}{110}}, \bibinfo{pages}{260601}}
  (\href{http://dx.doi.org/10.1103/PhysRevLett.110.260601}{\bibinfo{year}{2013}}).

\bibitem{mierzejewski2016}
\bibinfo{author}{M.~Mierzejewski}, \bibinfo{author}{J.~Herbrych}, and
  \bibinfo{author}{P.~Prelov\ifmmode~\check{s}\else \v{s}\fi{}ek},
  \bibinfo{title}{Universal dynamics of density correlations at the transition
  to the many-body localized state},
  \bibinfo{journal}{\href{http://dx.doi.org/10.1103/PhysRevB.94.224207}{Phys.
  Rev. B}} \href{http://dx.doi.org/10.1103/PhysRevB.94.224207}{{\bf
  \bibinfo{volume}{94}}, \bibinfo{pages}{224207}}
  (\href{http://dx.doi.org/10.1103/PhysRevB.94.224207}{\bibinfo{year}{2016}}).

\bibitem{luschen17}
\bibinfo{author}{H.~P. L\"uschen}, \bibinfo{author}{P.~Bordia},
  \bibinfo{author}{S.~Scherg}, \bibinfo{author}{F.~Alet},
  \bibinfo{author}{E.~Altman}, \bibinfo{author}{U.~Schneider}, and
  \bibinfo{author}{I.~Bloch}, \bibinfo{title}{Observation of slow dynamics near
  the many-body localization transition in one-dimensional quasiperiodic
  systems},
  \bibinfo{journal}{\href{http://dx.doi.org/10.1103/PhysRevLett.119.260401}{Phys.
  Rev. Lett.}} \href{http://dx.doi.org/10.1103/PhysRevLett.119.260401}{{\bf
  \bibinfo{volume}{119}}, \bibinfo{pages}{260401}}
  (\href{http://dx.doi.org/10.1103/PhysRevLett.119.260401}{\bibinfo{year}{2017}}).

\bibitem{bordia2017_1}
\bibinfo{author}{P.~Bordia}, \bibinfo{author}{H.~L\"uschen},
  \bibinfo{author}{S.~Scherg}, \bibinfo{author}{S.~Gopalakrishnan},
  \bibinfo{author}{M.~Knap}, \bibinfo{author}{U.~Schneider}, and
  \bibinfo{author}{I.~Bloch}, \bibinfo{title}{Probing slow relaxation and
  many-body localization in two-dimensional quasiperiodic systems},
  \bibinfo{journal}{\href{http://dx.doi.org/10.1103/PhysRevX.7.041047}{Phys.
  Rev. X}} \href{http://dx.doi.org/10.1103/PhysRevX.7.041047}{{\bf
  \bibinfo{volume}{7}}, \bibinfo{pages}{041047}}
  (\href{http://dx.doi.org/10.1103/PhysRevX.7.041047}{\bibinfo{year}{2017}}).

\bibitem{serbyn2017}
\bibinfo{author}{M.~Serbyn}, \bibinfo{author}{Z.~Papi\ifmmode~\acute{c}\else
  \'{c}\fi{}}, and \bibinfo{author}{D.~A. Abanin}, \bibinfo{title}{Thouless
  energy and multifractality across the many-body localization transition},
  \bibinfo{journal}{\href{http://dx.doi.org/10.1103/PhysRevB.96.104201}{Phys.
  Rev. B}} \href{http://dx.doi.org/10.1103/PhysRevB.96.104201}{{\bf
  \bibinfo{volume}{96}}, \bibinfo{pages}{104201}}
  (\href{http://dx.doi.org/10.1103/PhysRevB.96.104201}{\bibinfo{year}{2017}}).

\bibitem{Bera2019}
\bibinfo{author}{F.~Weiner}, \bibinfo{author}{F.~Evers}, and
  \bibinfo{author}{S.~Bera}, \bibinfo{title}{Slow dynamics and strong
  finite-size effects in many-body localization with random and quasiperiodic
  potentials},
  \bibinfo{journal}{\href{http://dx.doi.org/10.1103/PhysRevB.100.104204}{Phys.
  Rev. B}} \href{http://dx.doi.org/10.1103/PhysRevB.100.104204}{{\bf
  \bibinfo{volume}{100}}, \bibinfo{pages}{104204}}
  (\href{http://dx.doi.org/10.1103/PhysRevB.100.104204}{\bibinfo{year}{2019}}).

\bibitem{chanda2020}
\bibinfo{author}{T.~Chanda}, \bibinfo{author}{P.~Sierant}, and
  \bibinfo{author}{J.~Zakrzewski}, \bibinfo{title}{Time dynamics with matrix
  product states: Many-body localization transition of large systems
  revisited},
  \bibinfo{journal}{\href{http://dx.doi.org/10.1103/PhysRevB.101.035148}{Phys.
  Rev. B}} \href{http://dx.doi.org/10.1103/PhysRevB.101.035148}{{\bf
  \bibinfo{volume}{101}}, \bibinfo{pages}{035148}}
  (\href{http://dx.doi.org/10.1103/PhysRevB.101.035148}{\bibinfo{year}{2020}}).

\bibitem{berkelbach10}
\bibinfo{author}{T.~C. Berkelbach} and \bibinfo{author}{D.~R. Reichman},
  \bibinfo{title}{Conductivity of disordered quantum lattice models at infinite
  temperature: Many-body localization},
  \bibinfo{journal}{\href{http://dx.doi.org/10.1103/PhysRevB.81.224429}{Phys.
  Rev. B}} \href{http://dx.doi.org/10.1103/PhysRevB.81.224429}{{\bf
  \bibinfo{volume}{81}}, \bibinfo{pages}{224429}}
  (\href{http://dx.doi.org/10.1103/PhysRevB.81.224429}{\bibinfo{year}{2010}}).

\bibitem{lev15}
\bibinfo{author}{Y.~Bar~Lev}, \bibinfo{author}{G.~Cohen}, and
  \bibinfo{author}{D.~R. Reichman}, \bibinfo{title}{Absence of diffusion in an
  interacting system of spinless fermions on a one-dimensional disordered
  lattice},
  \bibinfo{journal}{\href{http://dx.doi.org/10.1103/PhysRevLett.114.100601}{Phys.
  Rev. Lett.}} \href{http://dx.doi.org/10.1103/PhysRevLett.114.100601}{{\bf
  \bibinfo{volume}{114}}, \bibinfo{pages}{100601}}
  (\href{http://dx.doi.org/10.1103/PhysRevLett.114.100601}{\bibinfo{year}{2015}}).

\bibitem{barisic16}
\bibinfo{author}{O.~S. Bari\ifmmode \check{s}\else
  \v{s}\fi{}i\ifmmode~\acute{c}\else \'{c}\fi{}}, \bibinfo{author}{J.~Kokalj},
  \bibinfo{author}{I.~Balog}, and
  \bibinfo{author}{P.~Prelov\ifmmode~\check{s}\else \v{s}\fi{}ek},
  \bibinfo{title}{Dynamical conductivity and its fluctuations along the
  crossover to many-body localization},
  \bibinfo{journal}{\href{http://dx.doi.org/10.1103/PhysRevB.94.045126}{Phys.
  Rev. B}} \href{http://dx.doi.org/10.1103/PhysRevB.94.045126}{{\bf
  \bibinfo{volume}{94}}, \bibinfo{pages}{045126}}
  (\href{http://dx.doi.org/10.1103/PhysRevB.94.045126}{\bibinfo{year}{2016}}).

\bibitem{steinigeweg16}
\bibinfo{author}{R.~Steinigeweg}, \bibinfo{author}{J.~Herbrych},
  \bibinfo{author}{F.~Pollmann}, and \bibinfo{author}{W.~Brenig},
  \bibinfo{title}{Typicality approach to the optical conductivity in thermal
  and many-body localized phases},
  \bibinfo{journal}{\href{http://dx.doi.org/10.1103/PhysRevB.94.180401}{Phys.
  Rev. B}} \href{http://dx.doi.org/10.1103/PhysRevB.94.180401}{{\bf
  \bibinfo{volume}{94}}, \bibinfo{pages}{180401}}
  (\href{http://dx.doi.org/10.1103/PhysRevB.94.180401}{\bibinfo{year}{2016}}).

\bibitem{bera2017}
\bibinfo{author}{S.~Bera}, \bibinfo{author}{G.~De~Tomasi},
  \bibinfo{author}{F.~Weiner}, and \bibinfo{author}{F.~Evers},
  \bibinfo{title}{Density propagator for many-body localization: Finite-size
  effects, transient subdiffusion, and exponential decay},
  \bibinfo{journal}{\href{http://dx.doi.org/10.1103/PhysRevLett.118.196801}{Phys.
  Rev. Lett.}} \href{http://dx.doi.org/10.1103/PhysRevLett.118.196801}{{\bf
  \bibinfo{volume}{118}}, \bibinfo{pages}{196801}}
  (\href{http://dx.doi.org/10.1103/PhysRevLett.118.196801}{\bibinfo{year}{2017}}).

\bibitem{prelovsek116}
\bibinfo{author}{P.~Prelov\ifmmode~\check{s}\else \v{s}\fi{}ek},
  \bibinfo{author}{M.~Mierzejewski}, \bibinfo{author}{O.~S. Bari\ifmmode
  \check{s}\else \v{s}\fi{}i\ifmmode~\acute{c}\else \'{c}\fi{}}, and
  \bibinfo{author}{J.~Herbrych}, \bibinfo{title}{Density correlations and
  transport in models of many‐body localization}, \bibinfo{journal}{Annalen
  der Physik} {\bf \bibinfo{volume}{529}}, \bibinfo{pages}{1600362}
  (\bibinfo{year}{2016}).

\bibitem{bardarson12}
\bibinfo{author}{J.~H. Bardarson}, \bibinfo{author}{F.~Pollmann}, and
  \bibinfo{author}{J.~E. Moore}, \bibinfo{title}{Unbounded growth of
  entanglement in models of many-body localization},
  \bibinfo{journal}{\href{http://dx.doi.org/10.1103/PhysRevLett.109.017202}{Phys.
  Rev. Lett.}} \href{http://dx.doi.org/10.1103/PhysRevLett.109.017202}{{\bf
  \bibinfo{volume}{109}}, \bibinfo{pages}{017202}}
  (\href{http://dx.doi.org/10.1103/PhysRevLett.109.017202}{\bibinfo{year}{2012}}).

\bibitem{kjall14}
\bibinfo{author}{J.~A. Kj{\"{a}}ll}, \bibinfo{author}{J.~H. Bardarson}, and
  \bibinfo{author}{F.~Pollmann}, \bibinfo{title}{Many-body localization in a
  disordered quantum {Ising} chain},
  \bibinfo{journal}{\href{http://dx.doi.org/10.1103/PhysRevLett.113.107204}{Phys.
  Rev. Lett.}} \href{http://dx.doi.org/10.1103/PhysRevLett.113.107204}{{\bf
  \bibinfo{volume}{113}}, \bibinfo{pages}{107204}}
  (\href{http://dx.doi.org/10.1103/PhysRevLett.113.107204}{\bibinfo{year}{2014}}).

\bibitem{baygan15}
\bibinfo{author}{E.~Baygan}, \bibinfo{author}{S.~P. Lim}, and
  \bibinfo{author}{D.~N. Sheng}, \bibinfo{title}{{Many-body localization and
  mobility edge in a disordered spin - 1/2 Heisenberg ladder}},
  \bibinfo{journal}{\href{http://dx.doi.org/10.1103/PhysRevB.92.195153}{Phys.
  Rev. B}} \href{http://dx.doi.org/10.1103/PhysRevB.92.195153}{{\bf
  \bibinfo{volume}{92}}, \bibinfo{pages}{195153}}
  (\href{http://dx.doi.org/10.1103/PhysRevB.92.195153}{\bibinfo{year}{2015}}).

\bibitem{pietracaprina_parisi_17}
\bibinfo{author}{F.~Pietracaprina}, \bibinfo{author}{G.~Parisi},
  \bibinfo{author}{A.~Mariano}, \bibinfo{author}{S.~Pascazio}, and
  \bibinfo{author}{A.~Scardicchio}, \bibinfo{title}{Entanglement critical
  length at the many-body localization transition},
  \bibinfo{journal}{\href{http://dx.doi.org/10.1088/1742-5468/aa9338}{J. Stat.
  Mech.}} \href{http://dx.doi.org/10.1088/1742-5468/aa9338}{{\bf
  \bibinfo{volume}{{\rm (2017)}}}, \bibinfo{pages}{113102}}
  (\href{http://dx.doi.org/10.1088/1742-5468/aa9338}{\bibinfo{year}{2017}}).

\bibitem{zakrzewski16}
\bibinfo{author}{P.~Sierant}, \bibinfo{author}{D.~Delande}, and
  \bibinfo{author}{J.~Zakrzewski}, \bibinfo{title}{Many-body localization due
  to random interactions},
  \bibinfo{journal}{\href{http://dx.doi.org/10.1103/PhysRevA.95.021601}{Phys.
  Rev. A}} \href{http://dx.doi.org/10.1103/PhysRevA.95.021601}{{\bf
  \bibinfo{volume}{95}}, \bibinfo{pages}{021601}}
  (\href{http://dx.doi.org/10.1103/PhysRevA.95.021601}{\bibinfo{year}{2017}}).

\bibitem{sierant_zakrzewski_19}
\bibinfo{author}{P.~Sierant} and \bibinfo{author}{J.~Zakrzewski},
  \bibinfo{title}{Level statistics across the many-body localization
  transition},
  \bibinfo{journal}{\href{http://dx.doi.org/10.1103/PhysRevB.99.104205}{Phys.
  Rev. B}} \href{http://dx.doi.org/10.1103/PhysRevB.99.104205}{{\bf
  \bibinfo{volume}{99}}, \bibinfo{pages}{104205}}
  (\href{http://dx.doi.org/10.1103/PhysRevB.99.104205}{\bibinfo{year}{2019}}).

\bibitem{kieferemmanouilidis_unanyan_20}
\bibinfo{author}{M.~Kiefer-Emmanouilidis}, \bibinfo{author}{R.~Unanyan},
  \bibinfo{author}{M.~Fleischhauer}, and \bibinfo{author}{J.~Sirker},
  \bibinfo{title}{Evidence for unbounded growth of the number entropy in
  many-body localized phases},
  \bibinfo{journal}{\href{http://dx.doi.org/10.1103/PhysRevLett.124.243601}{Phys.
  Rev. Lett.}} \href{http://dx.doi.org/10.1103/PhysRevLett.124.243601}{{\bf
  \bibinfo{volume}{124}}, \bibinfo{pages}{243601}}
  (\href{http://dx.doi.org/10.1103/PhysRevLett.124.243601}{\bibinfo{year}{2020}}).

\bibitem{sels2020}
\bibinfo{author}{D.~Sels} and \bibinfo{author}{A.~Polkovnikov},
  \bibinfo{title}{Dynamical obstruction to localization in a disordered spin
  chain},
  \bibinfo{journal}{\href{http://dx.doi.org/10.1103/PhysRevE.104.054105}{Phys.
  Rev. E}} \href{http://dx.doi.org/10.1103/PhysRevE.104.054105}{{\bf
  \bibinfo{volume}{104}}, \bibinfo{pages}{054105}}
  (\href{http://dx.doi.org/10.1103/PhysRevE.104.054105}{\bibinfo{year}{2021}}).

\bibitem{kieferemmanouilidis_unanyan_21}
\bibinfo{author}{M.~Kiefer-Emmanouilidis}, \bibinfo{author}{R.~Unanyan},
  \bibinfo{author}{M.~Fleischhauer}, and \bibinfo{author}{J.~Sirker},
  \bibinfo{title}{Slow delocalization of particles in many-body localized
  phases},
  \bibinfo{journal}{\href{http://dx.doi.org/10.1103/PhysRevB.103.024203}{Phys.
  Rev. B}} \href{http://dx.doi.org/10.1103/PhysRevB.103.024203}{{\bf
  \bibinfo{volume}{103}}, \bibinfo{pages}{024203}}
  (\href{http://dx.doi.org/10.1103/PhysRevB.103.024203}{\bibinfo{year}{2021}}).

\bibitem{leblond_sels_21}
\bibinfo{author}{T.~LeBlond}, \bibinfo{author}{D.~Sels},
  \bibinfo{author}{A.~Polkovnikov}, and \bibinfo{author}{M.~Rigol},
  \bibinfo{title}{Universality in the onset of quantum chaos in many-body
  systems},
  \bibinfo{journal}{\href{http://dx.doi.org/10.1103/PhysRevB.104.L201117}{Phys.
  Rev. B}} \href{http://dx.doi.org/10.1103/PhysRevB.104.L201117}{{\bf
  \bibinfo{volume}{104}}, \bibinfo{pages}{L201117}}
  (\href{http://dx.doi.org/10.1103/PhysRevB.104.L201117}{\bibinfo{year}{2021}}).

\bibitem{vidmar2021}
\bibinfo{author}{L.~Vidmar}, \bibinfo{author}{B.~Krajewski},
  \bibinfo{author}{J.~Bon\ifmmode~\check{c}\else \v{c}\fi{}a}, and
  \bibinfo{author}{M.~Mierzejewski}, \bibinfo{title}{Phenomenology of spectral
  functions in disordered spin chains at infinite temperature},
  \bibinfo{journal}{\href{http://dx.doi.org/10.1103/PhysRevLett.127.230603}{Phys.
  Rev. Lett.}} \href{http://dx.doi.org/10.1103/PhysRevLett.127.230603}{{\bf
  \bibinfo{volume}{127}}, \bibinfo{pages}{230603}}
  (\href{http://dx.doi.org/10.1103/PhysRevLett.127.230603}{\bibinfo{year}{2021}}).

\bibitem{Sels_dilute_2021}
\bibinfo{author}{D.~Sels} and \bibinfo{author}{A.~Polkovnikov},
  \bibinfo{title}{Thermalization of dilute impurities in one-dimensional spin
  chains},
  \bibinfo{journal}{\href{http://dx.doi.org/10.1103/PhysRevX.13.011041}{Phys.
  Rev. X}} \href{http://dx.doi.org/10.1103/PhysRevX.13.011041}{{\bf
  \bibinfo{volume}{13}}, \bibinfo{pages}{011041}}
  (\href{http://dx.doi.org/10.1103/PhysRevX.13.011041}{\bibinfo{year}{2023}}).

\bibitem{Sels_2022}
\bibinfo{author}{D.~Sels}, \bibinfo{title}{Bath-induced delocalization in
  interacting disordered spin chains},
  \bibinfo{journal}{\href{http://dx.doi.org/10.1103/PhysRevB.106.L020202}{Phys.
  Rev. B}} \href{http://dx.doi.org/10.1103/PhysRevB.106.L020202}{{\bf
  \bibinfo{volume}{106}}, \bibinfo{pages}{L020202}}
  (\href{http://dx.doi.org/10.1103/PhysRevB.106.L020202}{\bibinfo{year}{2022}}).

\bibitem{evers2023}
\bibinfo{author}{F.~Evers} and \bibinfo{author}{S.~Bera}, \bibinfo{title}{The
  internal clock of many-body (de-)localization},
  \href{https://arxiv.org/abs/2302.11384}{\bibinfo{howpublished}{arXiv:2302.11384}}
  (\bibinfo{year}{2023}).

\bibitem{sierant2022}
\bibinfo{author}{P.~Sierant} and \bibinfo{author}{J.~Zakrzewski},
  \bibinfo{title}{Challenges to observation of many-body localization},
  \bibinfo{journal}{\href{http://dx.doi.org/10.1103/PhysRevB.105.224203}{Phys.
  Rev. B}} \href{http://dx.doi.org/10.1103/PhysRevB.105.224203}{{\bf
  \bibinfo{volume}{105}}, \bibinfo{pages}{224203}}
  (\href{http://dx.doi.org/10.1103/PhysRevB.105.224203}{\bibinfo{year}{2022}}).

\bibitem{Morningstar2022}
\bibinfo{author}{A.~Morningstar}, \bibinfo{author}{L.~Colmenarez},
  \bibinfo{author}{V.~Khemani}, \bibinfo{author}{D.~J. Luitz}, and
  \bibinfo{author}{D.~A. Huse}, \bibinfo{title}{Avalanches and many-body
  resonances in many-body localized systems},
  \bibinfo{journal}{\href{http://dx.doi.org/10.1103/PhysRevB.105.174205}{Phys.
  Rev. B}} \href{http://dx.doi.org/10.1103/PhysRevB.105.174205}{{\bf
  \bibinfo{volume}{105}}, \bibinfo{pages}{174205}}
  (\href{http://dx.doi.org/10.1103/PhysRevB.105.174205}{\bibinfo{year}{2022}}).

\bibitem{abanin2019}
\bibinfo{author}{D.~A. Abanin}, \bibinfo{author}{E.~Altman},
  \bibinfo{author}{I.~Bloch}, and \bibinfo{author}{M.~Serbyn},
  \bibinfo{title}{Colloquium: Many-body localization, thermalization, and
  entanglement},
  \bibinfo{journal}{\href{http://dx.doi.org/10.1103/RevModPhys.91.021001}{Rev.
  Mod. Phys.}} \href{http://dx.doi.org/10.1103/RevModPhys.91.021001}{{\bf
  \bibinfo{volume}{91}}, \bibinfo{pages}{021001}}
  (\href{http://dx.doi.org/10.1103/RevModPhys.91.021001}{\bibinfo{year}{2019}}).

\bibitem{Panda2020}
\bibinfo{author}{R.~K. Panda}, \bibinfo{author}{A.~Scardicchio},
  \bibinfo{author}{M.~Schulz}, \bibinfo{author}{S.~R. Taylor}, and
  \bibinfo{author}{M.~{\v{Z}}nidari{\v{c}}}, \bibinfo{title}{Can we study the
  many-body localisation transition?},
  \bibinfo{journal}{\href{http://dx.doi.org/10.1209/0295-5075/128/67003}{{EPL}
  (Europhysics Letters)}}
  \href{http://dx.doi.org/10.1209/0295-5075/128/67003}{{\bf
  \bibinfo{volume}{128}}, \bibinfo{pages}{67003}}
  (\href{http://dx.doi.org/10.1209/0295-5075/128/67003}{\bibinfo{year}{2020}}).

\bibitem{abanin_bardarson_21}
\bibinfo{author}{D.~Abanin}, \bibinfo{author}{J.~Bardarson},
  \bibinfo{author}{G.~{De Tomasi}}, \bibinfo{author}{S.~Gopalakrishnan},
  \bibinfo{author}{V.~Khemani}, \bibinfo{author}{S.~Parameswaran},
  \bibinfo{author}{F.~Pollmann}, \bibinfo{author}{A.~Potter},
  \bibinfo{author}{M.~Serbyn}, and \bibinfo{author}{R.~Vasseur},
  \bibinfo{title}{{Distinguishing localization from chaos: Challenges in
  finite-size systems}},
  \bibinfo{journal}{\href{http://dx.doi.org/https://doi.org/10.1016/j.aop.2021.168415}{Ann.
  Phys.}}
  \href{http://dx.doi.org/https://doi.org/10.1016/j.aop.2021.168415}{{\bf
  \bibinfo{volume}{427}}, \bibinfo{pages}{168415}}
  (\href{http://dx.doi.org/https://doi.org/10.1016/j.aop.2021.168415}{\bibinfo{year}{2021}}).

\bibitem{crowley_chandran_22}
\bibinfo{author}{P.~J.~D. Crowley} and \bibinfo{author}{A.~Chandran},
  \bibinfo{title}{{A constructive theory of the numerically accessible
  many-body localized to thermal crossover}},
  \bibinfo{journal}{\href{http://dx.doi.org/10.21468/SciPostPhys.12.6.201}{SciPost
  Phys.}} \href{http://dx.doi.org/10.21468/SciPostPhys.12.6.201}{{\bf
  \bibinfo{volume}{12}}, \bibinfo{pages}{201}}
  (\href{http://dx.doi.org/10.21468/SciPostPhys.12.6.201}{\bibinfo{year}{2022}}).

\bibitem{bolther_kehrein_22}
\bibinfo{author}{N.~B\"olter} and \bibinfo{author}{S.~Kehrein},
  \bibinfo{title}{{Scrambling and many-body localization in the XXZ chain}},
  \bibinfo{journal}{\href{http://dx.doi.org/10.1103/PhysRevB.105.104202}{Phys.
  Rev. B}} \href{http://dx.doi.org/10.1103/PhysRevB.105.104202}{{\bf
  \bibinfo{volume}{105}}, \bibinfo{pages}{104202}}
  (\href{http://dx.doi.org/10.1103/PhysRevB.105.104202}{\bibinfo{year}{2022}}).

\bibitem{deluca13}
\bibinfo{author}{A.~De~Luca} and \bibinfo{author}{A.~Scardicchio},
  \bibinfo{title}{Ergodicity breaking in a model showing many-body
  localization},
  \bibinfo{journal}{\href{http://dx.doi.org/10.1209/0295-5075/101/37003}{EPL
  (Europhysics Letters)}}
  \href{http://dx.doi.org/10.1209/0295-5075/101/37003}{{\bf
  \bibinfo{volume}{101}}, \bibinfo{pages}{37003}}
  (\href{http://dx.doi.org/10.1209/0295-5075/101/37003}{\bibinfo{year}{2013}}).

\bibitem{krajewski_vidmar_22}
\bibinfo{author}{B.~Krajewski}, \bibinfo{author}{L.~Vidmar},
  \bibinfo{author}{J.~Bon\ifmmode~\check{c}\else \v{c}\fi{}a}, and
  \bibinfo{author}{M.~Mierzejewski}, \bibinfo{title}{Restoring ergodicity in a
  strongly disordered interacting chain},
  \bibinfo{journal}{\href{http://dx.doi.org/10.1103/PhysRevLett.129.260601}{Phys.
  Rev. Lett.}} \href{http://dx.doi.org/10.1103/PhysRevLett.129.260601}{{\bf
  \bibinfo{volume}{129}}, \bibinfo{pages}{260601}}
  (\href{http://dx.doi.org/10.1103/PhysRevLett.129.260601}{\bibinfo{year}{2022}}).

\bibitem{deutsch_91}
\bibinfo{author}{J.~M. Deutsch}, \bibinfo{title}{Quantum statistical mechanics
  in a closed system},
  \bibinfo{journal}{\href{http://dx.doi.org/10.1103/PhysRevA.43.2046}{Phys.
  Rev. A}} \href{http://dx.doi.org/10.1103/PhysRevA.43.2046}{{\bf
  \bibinfo{volume}{43}}, \bibinfo{pages}{2046}}
  (\href{http://dx.doi.org/10.1103/PhysRevA.43.2046}{\bibinfo{year}{1991}}).

\bibitem{srednicki_94}
\bibinfo{author}{M.~Srednicki}, \bibinfo{title}{Chaos and quantum
  thermalization},
  \bibinfo{journal}{\href{http://dx.doi.org/10.1103/PhysRevE.50.888}{Phys. Rev.
  E}} \href{http://dx.doi.org/10.1103/PhysRevE.50.888}{{\bf
  \bibinfo{volume}{50}}, \bibinfo{pages}{888}}
  (\href{http://dx.doi.org/10.1103/PhysRevE.50.888}{\bibinfo{year}{1994}}).

\bibitem{rigol_dunjko_08}
\bibinfo{author}{M.~Rigol}, \bibinfo{author}{V.~Dunjko}, and
  \bibinfo{author}{M.~Olshanii}, \bibinfo{title}{Thermalization and its
  mechanism for generic isolated quantum systems},
  \bibinfo{journal}{\href{http://dx.doi.org/10.1038/nature06838}{Nature
  (London)}} \href{http://dx.doi.org/10.1038/nature06838}{{\bf
  \bibinfo{volume}{452}}, \bibinfo{pages}{854}}
  (\href{http://dx.doi.org/10.1038/nature06838}{\bibinfo{year}{2008}}).

\bibitem{dalessio_kafri_16}
\bibinfo{author}{L.~D'Alessio}, \bibinfo{author}{Y.~Kafri},
  \bibinfo{author}{A.~Polkovnikov}, and \bibinfo{author}{M.~Rigol},
  \bibinfo{title}{From quantum chaos and eigenstate thermalization to
  statistical mechanics and thermodynamics},
  \bibinfo{journal}{\href{http://dx.doi.org/10.1080/00018732.2016.1198134}{Adv.
  Phys.}} \href{http://dx.doi.org/10.1080/00018732.2016.1198134}{{\bf
  \bibinfo{volume}{65}}, \bibinfo{pages}{239}}
  (\href{http://dx.doi.org/10.1080/00018732.2016.1198134}{\bibinfo{year}{2016}}).

\bibitem{atas2013}
\bibinfo{author}{Y.~Y. Atas}, \bibinfo{author}{E.~Bogomolny},
  \bibinfo{author}{O.~Giraud}, and \bibinfo{author}{G.~Roux},
  \bibinfo{title}{Distribution of the ratio of consecutive level spacings in
  random matrix ensembles},
  \bibinfo{journal}{\href{http://dx.doi.org/10.1103/PhysRevLett.110.084101}{Phys.
  Rev. Lett.}} \href{http://dx.doi.org/10.1103/PhysRevLett.110.084101}{{\bf
  \bibinfo{volume}{110}}, \bibinfo{pages}{084101}}
  (\href{http://dx.doi.org/10.1103/PhysRevLett.110.084101}{\bibinfo{year}{2013}}).

\bibitem{Husex2017}
\bibinfo{author}{V.~Khemani}, \bibinfo{author}{S.~P. Lim},
  \bibinfo{author}{D.~N. Sheng}, and \bibinfo{author}{D.~A. Huse},
  \bibinfo{title}{Critical properties of the many-body localization
  transition},
  \bibinfo{journal}{\href{http://dx.doi.org/10.1103/PhysRevX.7.021013}{Phys.
  Rev. X}} \href{http://dx.doi.org/10.1103/PhysRevX.7.021013}{{\bf
  \bibinfo{volume}{7}}, \bibinfo{pages}{021013}}
  (\href{http://dx.doi.org/10.1103/PhysRevX.7.021013}{\bibinfo{year}{2017}}).

\bibitem{sirker14}
\bibinfo{author}{F.~Andraschko}, \bibinfo{author}{T.~Enss}, and
  \bibinfo{author}{J.~Sirker}, \bibinfo{title}{Purification and many-body
  localization in cold atomic gases},
  \bibinfo{journal}{\href{http://dx.doi.org/10.1103/PhysRevLett.113.217201}{Phys.
  Rev. Lett.}} \href{http://dx.doi.org/10.1103/PhysRevLett.113.217201}{{\bf
  \bibinfo{volume}{113}}, \bibinfo{pages}{217201}}
  (\href{http://dx.doi.org/10.1103/PhysRevLett.113.217201}{\bibinfo{year}{2014}}).

\bibitem{Kim_strong2014}
\bibinfo{author}{H.~Kim}, \bibinfo{author}{T.~N. Ikeda}, and
  \bibinfo{author}{D.~A. Huse}, \bibinfo{title}{Testing whether all eigenstates
  obey the eigenstate thermalization hypothesis},
  \bibinfo{journal}{\href{http://dx.doi.org/10.1103/PhysRevE.90.052105}{Phys.
  Rev. E}} \href{http://dx.doi.org/10.1103/PhysRevE.90.052105}{{\bf
  \bibinfo{volume}{90}}, \bibinfo{pages}{052105}}
  (\href{http://dx.doi.org/10.1103/PhysRevE.90.052105}{\bibinfo{year}{2014}}).

\bibitem{Mondaini2016}
\bibinfo{author}{R.~Mondaini}, \bibinfo{author}{K.~R. Fratus},
  \bibinfo{author}{M.~Srednicki}, and \bibinfo{author}{M.~Rigol},
  \bibinfo{title}{{Eigenstate thermalization in the two-dimensional transverse
  field Ising model}},
  \bibinfo{journal}{\href{http://dx.doi.org/10.1103/PhysRevE.93.032104}{Phys.
  Rev. E}} \href{http://dx.doi.org/10.1103/PhysRevE.93.032104}{{\bf
  \bibinfo{volume}{93}}, \bibinfo{pages}{032104}}
  (\href{http://dx.doi.org/10.1103/PhysRevE.93.032104}{\bibinfo{year}{2016}}).

\bibitem{jansen_stolpp_19}
\bibinfo{author}{D.~Jansen}, \bibinfo{author}{J.~Stolpp},
  \bibinfo{author}{L.~Vidmar}, and \bibinfo{author}{F.~Heidrich-Meisner},
  \bibinfo{title}{{Eigenstate thermalization and quantum chaos in the Holstein
  polaron model}},
  \bibinfo{journal}{\href{http://dx.doi.org/10.1103/PhysRevB.99.155130}{Phys.
  Rev. B}} \href{http://dx.doi.org/10.1103/PhysRevB.99.155130}{{\bf
  \bibinfo{volume}{99}}, \bibinfo{pages}{155130}}
  (\href{http://dx.doi.org/10.1103/PhysRevB.99.155130}{\bibinfo{year}{2019}}).

\bibitem{mierzejewski2020}
\bibinfo{author}{M.~Mierzejewski}, \bibinfo{author}{M.~\ifmmode~\acute{S}\else
  \'{S}\fi{}roda}, \bibinfo{author}{J.~Herbrych}, and
  \bibinfo{author}{P.~Prelov\ifmmode~\check{s}\else \v{s}\fi{}ek},
  \bibinfo{title}{Resistivity and its fluctuations in disordered many-body
  systems: From chains to planes},
  \bibinfo{journal}{\href{http://dx.doi.org/10.1103/PhysRevB.102.161111}{Phys.
  Rev. B}} \href{http://dx.doi.org/10.1103/PhysRevB.102.161111}{{\bf
  \bibinfo{volume}{102}}, \bibinfo{pages}{161111}}
  (\href{http://dx.doi.org/10.1103/PhysRevB.102.161111}{\bibinfo{year}{2020}}).

\bibitem{krajewski2022}
\bibinfo{author}{B.~Krajewski}, \bibinfo{author}{M.~Mierzejewski}, and
  \bibinfo{author}{J.~Bon\ifmmode~\check{c}\else \v{c}\fi{}a},
  \bibinfo{title}{Modeling sample-to-sample fluctuations of the gap ratio in
  finite disordered spin chains},
  \bibinfo{journal}{\href{http://dx.doi.org/10.1103/PhysRevB.106.014201}{Phys.
  Rev. B}} \href{http://dx.doi.org/10.1103/PhysRevB.106.014201}{{\bf
  \bibinfo{volume}{106}}, \bibinfo{pages}{014201}}
  (\href{http://dx.doi.org/10.1103/PhysRevB.106.014201}{\bibinfo{year}{2022}}).

\end{thebibliography}
 
\end{document}